\documentclass[aps,prd,superscriptaddress,notitlepage,twocolumn,bibnotes
]{revtex4-1}
\usepackage{graphicx,amsmath,mathtools,amsfonts}
\usepackage{bm}
\usepackage[normalem]{ulem}
\usepackage[usenames, dvipsnames]{xcolor}
\usepackage{multirow}
\usepackage{tabularx}
\usepackage{soul} 
\usepackage{svg}  
\usepackage{hyperref}
\hypersetup{colorlinks=true, linkcolor=blue, citecolor=blue, urlcolor=blue}
\usepackage{comment}
\usepackage{ulem}

\begin{document}

\title{
Pressure and doping effects on the \\
electronic structure and magnetism of the single-layer nickelate La$_2$NiO$_4$
}

\author{J. B. de Vaulx}
\affiliation{Univ. Grenoble Alpes, CNRS, Grenoble INP, Institut Néel, 25 Rue des Martyrs, 38042, Grenoble, France}
\author{F. Bernardini}
\affiliation{Dipartimento di Fisica, Universit\`a di Cagliari, IT-09042 Monserrato, Italy}
\author{V. Olevano}
\affiliation{Univ. Grenoble Alpes, CNRS, Grenoble INP, Institut Néel, 25 Rue des Martyrs, 38042, Grenoble, France}
\author{Q. N. Meier}
\affiliation{Univ. Grenoble Alpes, CNRS, Grenoble INP, Institut Néel, 25 Rue des Martyrs, 38042, Grenoble, France}
\author{A. Cano}
\email{andres.cano@neel.cnrs.fr}
\affiliation{Univ. Grenoble Alpes, CNRS, Grenoble INP, Institut Néel, 25 Rue des Martyrs, 38042, Grenoble, France}

\date{\today}

\begin{abstract}  
La$_2$NiO$_4$ is a prototypical member of the Ruddlesden-Popper nickelate series that offers a valuable reference point for elucidating the key ingredients behind the intriguing properties of these systems. 
However, the structural and electronic properties of La$_2$NiO$_4$ under pressure and doping remain surprisingly underexplored. 
Here, we investigate these properties using density-functional-theory calculations. 
We find that its tetragonal $I4/mmm$ structure can be stabilized, not only under {pressure,} but also at ambient pressure via the partial substitution of La with Ba. 
{In both cases, we find a pronounced magnetostructural interplay that manifests, in particular, as anomalies in the lattice-parameter evolution with composition, deviating from Vegard's law.} Moreover, we show that the combined effects of Ba substitution and pressure leads to qualitative changes in the electronic structure towards the formal $d^{7.5}$ configuration of the superconducting bilayer nickelates. 
Further, while La$_2$NiO$_4$ can undergo a insulator-metal transition with pressure retaining G-type antiferromagnetic order, La$_{1.5}$Ba$_{0.5}$NiO$_4$ exhibits metallic behavior with an enhanced competition between different magnetic states. Our results thus offer new insights into the interplay of structure, doping, and magnetism across the Ruddlesden-Popper nickelate series.
\end{abstract}

\maketitle

{Ruddlesden-Popper nickelates serve both as model systems for studying intertwined charge and magnetic orders \cite{lorenzo94}, including altermagnetism \cite{bernardini25}, as well as functional materials with redox flexibility and mixed electronic-ionic conductivity for example that make them attractive for applications in electrocatalysis, solid oxide fuel cells and next-generation batteries (see  e.g. \cite{Wu2024,bassat25,Wissel2021}). 
In addition, they have recently emerged as a compelling new family of high-$T_c$ superconductors, with bulk samples exhibiting superconducting transition onsets up to 92~K under pressure \cite{sun23,li24,wang24,zhang25}. Besides,} 
superconductivity has also been reported in thin films under epitaxial strain \cite{Hwang24,Zhuoyu24}. Similarly to their infinite-layer counterparts \cite{arita19,held23,meier24}, the emergence of superconductivity in these systems is difficult to reconcile with the conventional electron-phonon mechanism \cite{Ouyang2024,You2025}. 
However, these nickelates appear as a distinct class of unconventional superconductors featuring a $d^{7.5}$ electronic configuration with specific multiband features. La$_3$Ni$_2$O$_7$, in particular, has been predicted to support $s_\pm$-wave pairing from its correlated Ni-$d_{z^2}$ states \cite{kuroki17-prb} (see also \cite{kuroki24-prl}). However, alongside the $s_\pm$-wave state, $d$-wave pairing has also been proposed, which depends on specific details of the electronic structure and calls for further clarification (see e.g. 
\cite{lechermann23-prb,wu24-prl,savrasov24-prb,dagotto24,xia25-ncomm,xu25,gao25,maier25,kuroki25,ryee25,cano25ffJJ}). 

Progress in the field, however, requires a broader consideration of the entire Ruddlesden-Popper nickelate family, including their magnetic properties. Early magnetic susceptibility measurements on nominal La$_3$Ni$_2$O$_7$ samples, for example, indicated very low superconducting volume fractions \cite{zhou2024filamentary}, raising concerns about secondary phases and structural polymorphs arising from intergrowth. Partial substitution of La with Pr or Sm has been shown to stabilize a nearly pure bilayer phase \cite{wang24,zhang25}, yet completely eliminating intergrowth with other members of the series remains an experimental challenge. Furthermore, neutron scattering measurements have recently revealed long-range magnetic order in both La$_3$Ni$_2$O$_7$ and La$_2$PrNi$_2$O$_7$ at ambient pressure \cite{plokhikh25}. In this context, the single-layer ($n=1$) nickelates warrant renewed attention, not only to address synthesis and phase-purity issues, but also to serve as a reference point for a comprehensive understanding the electronic and magnetic properties across the Ruddlesden-Popper nickelate series.

\begin{figure}[b!]
    \centering
\includegraphics[height=.275\textwidth]{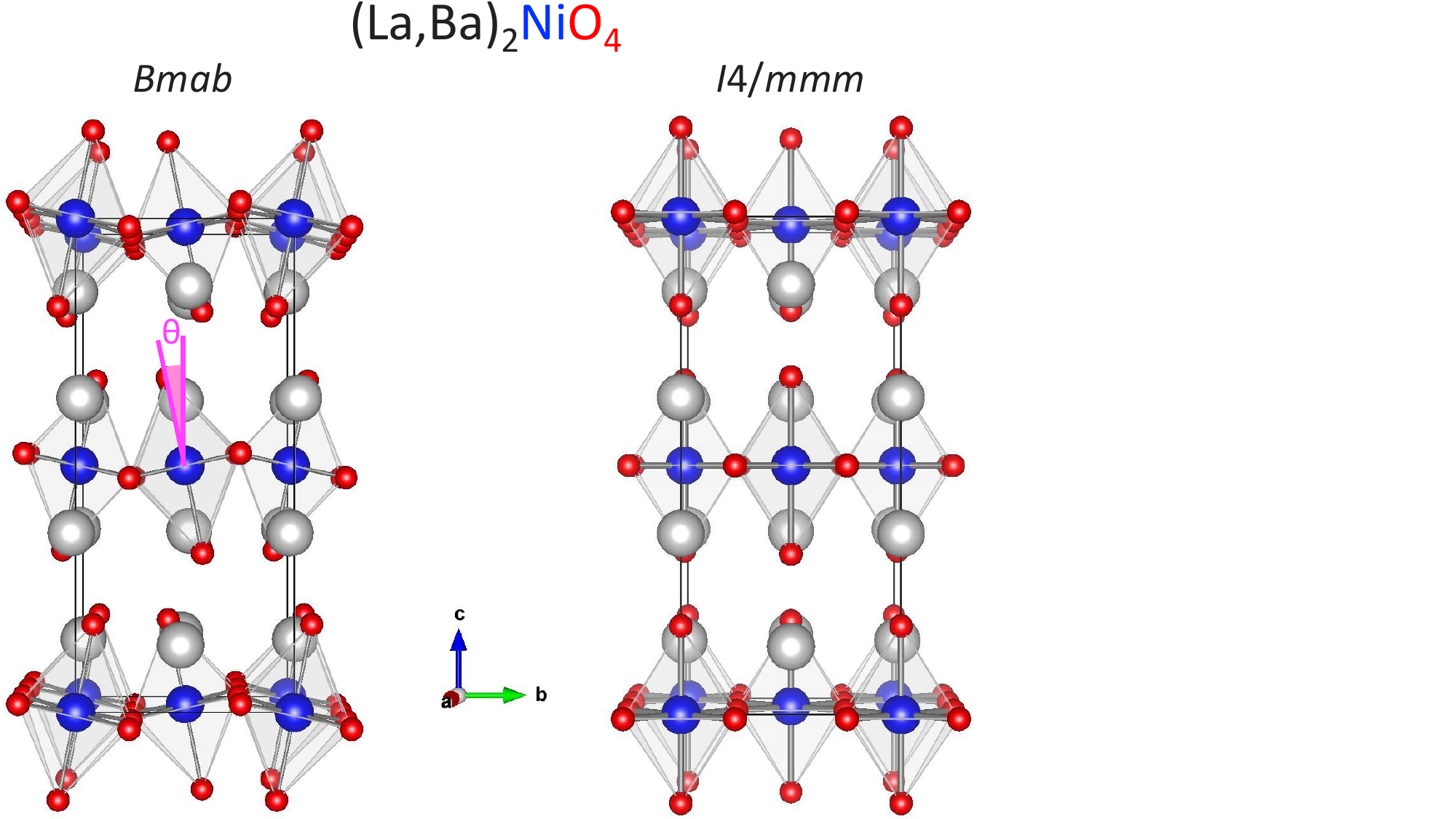}
    \caption{(a) Ball-and-stick model of the single-layer nickelates illustrating the orthorhombic ($Bmab$) and tetragonal ($I4/mmm$) structures{---with and without NiO$_6$ octahedra tilts---characteristic of these systems.}
}
\label{fig:SL}
\end{figure}

In this work, we investigate the properties of the single-layer nickelates as a function of pressure and composition.
Specifically, we use density-functional theory (DFT) to compute the structural phase diagram of La$_2$NiO$_4$ {(see Fig. \ref{fig:SL})}. 
Our calculations show that its crystal structure can be driven to a high-symmetry tetragonal phase by the application of pressure, which can be either physical, ``chemical'' or both.  
As an illustration, we consider partial substitution of La with Ba, which not only stabilizes the tetragonal phase but also formally dopes the system toward the $d^{7.5}$ configuration of the superconducting bilayer La$_3$Ni$_2$O$_7$. Further, we study the electronic structure and magnetic tendencies of (La,Ba)$_2$NiO$_4$, revealing both analogies and differences compared to the superconducting members of the Ruddlesden-Popper series. 
These results underscore the single-layer case as a valuable reference system for clarifying not only the emergence of magnetic order, but also 
{the possible} 
competition among distinct superconducting instabilities in the nickelates.

\vspace{.5ex}
\noindent{\bf Methods.} \\
\noindent The structural optimizations and non-spin polarized DFT calculations were performed using \textsc{Quantum Espresso} \cite{QE}. In these calculations, we considered the GGA approximation using the PBE form of the exchange-correlation functional with PseudoDojo pseudopotentials \cite{PBE,Dojo}. The calculations were converged using a $16\times 16 \times 16$ $k$-point mesh for the $I4/mmm$ structure and a $12\times 12 \times 12$ for the $Bmab$ one, with a 100 (400) Ry cutoff for wavefunction (density) and a Gaussian smearing of 0.01~Ry. 
For the spin-polarized calculations, we considered the previously optimized crystal structures and used the all-electron code {\sc{WIEN2k}} \cite{WIEN2k} with the LDA+U exchange-correlation functional \cite{LDA+U}. 
In these calculations, we used muffin-tin radii of 2.5, 2.0, and 1.60 a.u. for the La, Ni, and O atoms, respectively, and a plane-wave cutoff $R_{\rm MT}K_{\rm max}$ = 6.0. 
The integration over the Brillouin zone was done using a Monkhorst-Pack mesh of $10\times 10 \times 10$ and tetrahedron method \cite{tetrahedron} for the primitive cell and equivalent ones for the supercells when computing the magnetically ordered structures.
{Additional structural optimizations were performed for the G-type antiferromagnetic (G-AFM) configuration using the corresponding primitive cell and a $10\times 10 \times 10$ Monkhorst-Pack $k$-mesh with a 125 (500) Ry cutoff for the wavefunction (density) and a Gaussian smearing of 0.01~Ry. Structures were visualized with VESTA \cite{vesta}.
}

{
To address compositional effects, we focused on the La~$\to$~Ba substitution and employed the virtual crystal approximation (VCA). 
From the methodological point of view, this approach is particularly appropriate since i) the relevant electronic shells in La$^{3+}$ and Ba$^{2+}$ are the same \cite{cano20a} 
and ii) they are treated on the same footing using \textsc{Quantum Espresso} and the PeudoDojo. 
In addition, to the best of our knowledge, there is no experimental evidence of positional La/Ba ordering (see e.g. \cite{labani04-89,alonso90,yaremchenko23,schilling09}), further justifying the use of VCA. For La$_{1.5}$Ba$_{0.5}$NiO$_4$ in particular, we thus obtain $a = 3.887$~{\AA} and $c = 12.78$~{\AA} in the non-spin polarized case, in close agreement with the experimental values $a = 3.842$~{\AA} and $c = 12.86$~{\AA} \cite{labani04-89}. 
The reliability of VCA is further corroborated by the relaxed structure of LaBaNiO$_4$, for which VCA yields the lattice parameters $a = 3.942$~{\AA} and $c = 12.69$~{\AA}, essentially identical to the explicit La~$\to$~Ba substitution in the primitive $I4/mmm$ ($a = 3.942$~{\AA} and $c = 12.61$~{\AA}).
}

\begin{figure}[b!]
\includegraphics[width=.38\textwidth]{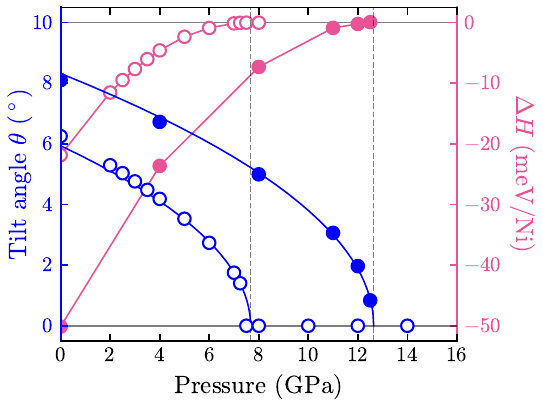}
\caption{
{Calculated tilt angle $\theta$ 
   and enthalpy $\Delta H$ as a function of pressure in La$_{2}$NiO$_4$. Empty and filled circles correspond to non-spin polarized calculations and G-AFM order respectively.
   In both cases, the supression of the tilt as a funcion of pressure scales as $\theta \propto |P-P_c|^{1/2}$ as indicated by the lines.} 
}
\label{fig:tilt.vs.p}
\end{figure}

\vspace{.5em}
\noindent{{\bf Pressure and compositional effects.}} \\
\noindent We begin by examining the structural phase diagram of La$_2$NiO$_4$ under pressure. 
At ambient pressure, {La$_2$NiO$_4$} adopts an orthorhombic structure with space group $Bmab$, characterized by tilted NiO$_6$ octahedra, as illustrated in Fig.~\ref{fig:SL}~\cite{rodriguez-carvajal91}. 
Symmetry-wise, this phase is a subgroup of the high-symmetry tetragonal $I4/mmm$ one also illustrated in Fig.~\ref{fig:SL} (where the octahedral tilting is directly related to the X$_3^+$ irreducible representation of the latter). 
Our {non-spin polarized} calculations reveal that applying pressure gradually suppresses the octahedral tilts as shown in Fig. \ref{fig:tilt.vs.p} {(empty circles)}.
The corresponding enthalpy difference, $\Delta H = \Delta E + P\Delta V$, is also plotted in Fig. \ref{fig:tilt.vs.p}. 
{
According to these results, there should be a continuous pressure-induced $Bmab \leftrightarrow I4/mmm$ phase transition in La$_2$NiO$_4$ at 8 GPa.
}

{
However, additional spin-polarized calculations for the experimental G-AFM configuration reveal a significant magnetostructural coupling in La$_2$NiO$_4$. 
In particular, the calculated lattice parameters at ambient pressure change from $(a,b,c)= (5.328, 5.465, 13.02)$~{\AA} in the non-spin polarized case to 
$(a, b, c) = (5.459, 5.649, 12.46)$ {\AA} in the presence of G-AFM order. 
The latter values are in better agreement with the experimental data (see e.g. \cite{rodriguez-carvajal91}).
Furthermore, the presence of magnetic order produces a hardening effect that shifts the calculated critical pressure of the structural transition to $\sim 12.5$~GPa (filled circles in Fig.~\ref{fig:tilt.vs.p}). 
Apart from this shift, we find that the overall behavior remains unchanged, in the sense that non-spin polarized and spin-polarized calculations cases both yield the Landau-theory scaling relation $\theta \propto |P-P_c|^{1/2}$ for the tilt angle. 
Interestingly, we find that the Ni magnetic moment does not vanish at $P_c$ (although it decreases slightly from 1.32~$ \mu_{\rm B}$ at ambient pressure to 1.23~$\mu_{\rm B}$).   
}

{
Further, the calculated Ni-O bond lengths show no significant change under pressure. 
Specifically, in the G-AFM configuration, the apical distance changes from 2.24 to 2.17~{\AA} only 
between ambient pressure and 12.5~GPa. 
The equatorial bond, in its turn, decreases from 1.97 to 1.89~{\AA}. To the best of our knowledge, 
experimental data are only available for non-stoichiometric La$_2$NiO$_{4+\delta}$ and Pr$_2$NiO$_{4+\delta}$ \cite{playford20}, where excess oxygen likely distorts the octahedra, precluding direct comparison. Nevertheless, both calculated and reported changes are small, indicating that the quasi-2D layered character 
is preserved across the pressure-driven structural transition.
}

\begin{figure}[t!]
\includegraphics[width=.45\textwidth]{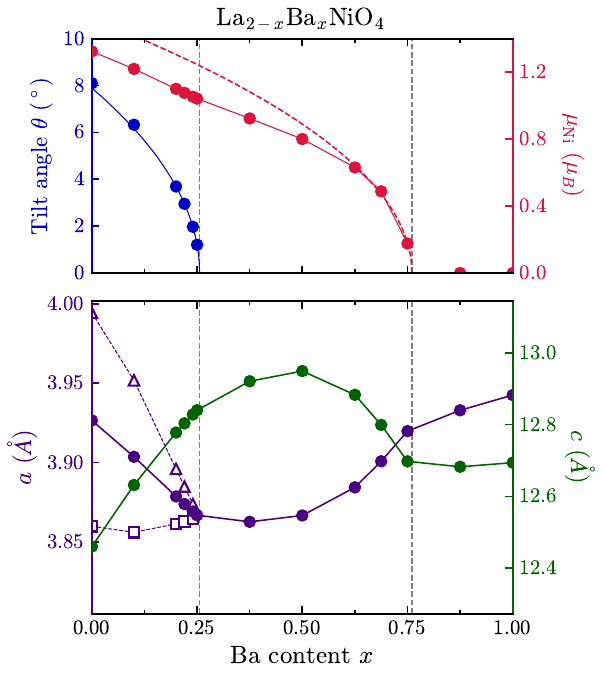}
\caption{
{(Top panel) Calculated tilt angle $\theta$ and Ni magnetic moment $\mu_\text{Ni}$ as a function of the Ba content $x$ in La$_{2-x}$Ba$_x$NiO$_4$ (G-AFM configuration). 
   Both the suppression of the tilt angle and the quenching of the Ni magnetic moment scale as $|x-x_c|^{1/2}$
   (blue line for $\theta$ and red dashed line for $\mu_\text{Ni}$). (Bottom panel) Calculated lattice parameters as a function of the Ba content $x$ in La$_{2-x}$Ba$_x$NiO$_4$. 
   Circles represent ``pseudo-tetragonal" or tetragonal lattice parameters ($a_\text{pseudo-t} = \sqrt{ab/2}$ or $a$ and $c$). 
   Squares and triangles correspond to the orthorhombic $a/\sqrt{2}$ and $b/\sqrt{2}$ respectively, whose difference correlates with the emergence of the tilt.  
}
}
\label{fig:m_ac.vs.x}
\end{figure}

{
We now turn to chemical pressure effects by considering the substitution of La
with Ba, whose larger ionic radius mimics the action of pressure. 
We find that Ba substitution in fact reproduces the effect hydrostatic pressure on the crystal structure, in the sense that it leads to the gradual suppression of the octahedral tilts according to a Landau-theory scaling $\theta \propto |x-x_{c,\theta}|^{1/2}$ with $x_{c,\theta} \sim 0.25$. 
This behavior is shown in the top panel of Figure~\ref{fig:m_ac.vs.x}, which correspond to spin-polarized calculations in the G-AFM configuration (blue circles and blue solid line for the $\theta \propto |x-x_{c,\theta}|^{1/2}$ scaling) 
\footnote{
The calculated structural phase diagram as a function of pressure and doping is shown in Figure~\ref{fig:PhD} for the non-spin polarized case.}. 
} 

{
In addition, we find that Ba substitution leads to the subsequent suppression of the Ni magnetic moment in the $G$-AFM configuration, which eventually vanishes following a similar scaling 
$\mu_{\rm Ni} \propto |x-x_{c,\mu_{\rm Ni}}|^{1/2}$ with $x_{c,\mu_{\rm Ni}} \sim 0.75$.
As under pressure, we thus find a composition range in which the system becomes tetragonal yet retaining $G$-AFM order (opposite to the temperature-driven behavior, where the tilts persist in the absence of magnetic order \cite{rodriguez-carvajal91}). 
This demonstrates that, despite the presence of sizable magnetostructural coupling, the structural and magnetic instabilities are fundamentally distinct and can be tuned independently. This further implies that, in the general temperature-pressure-composition phase space, there should be a point in which both these transitions coincide, giving rise to multiferroic critical behavior \cite{narayan19}. 
}

{
Further insight into the magnetostructural coupling is provided by the calculated lattice parameters as a function of Ba content (Fig.~\ref{fig:m_ac.vs.x}, bottom panel).
These exhibit a non-monotonic behavior, deviating from Vegard's law, that is in remarkable agreement with the experimental data 
(see \cite{labani04-89}, as well as \cite{yaremchenko23} and the references therein). 
By comparison with the tilt angle and the Ni magnetic moment, we find that this unusual trend is directly correlated with the structural and magnetic transitions.

To understand this correlation, it is convenient to take LaBaNiO$_4$ as the parent compound ($x=1$ in Fig. \ref{fig:m_ac.vs.x}). Thus, for decreasing Ba content, Vegard's law is initially obeyed in the absence of both octahedral tilts and magnetic order. This means that the free energy can be written as $F_0 = g_1 x' (u_{xx} + u_{yy})+ g_3 x' u_{zz} + {K \over 2}u_{ii}^2$, where $x' = 1 - x$ and $u_{ij}$ is the strain tensor expressed in the orthorhombic axes ($K$ is the bulk modulus and $g_i$ are constants). 
The diagonal components of $u_{ij}$ embody the relative variation of the lattice parameters which, in equilibrium, will then scale linearly with $x'$ (Vegard's law). 
However, additional couplings of the strain to the magnetic and structural order parameters $L$ and $Q$ are also allowed by symmetry: 
\begin{align}
F_L &= g_{1,L} L^2 (u_{xx}+ u_{yy}) + g_{3,L} L^2 u_{zz}, \\ 
F_Q &= g_{1,Q} Q^2 (u_{xx}-u_{yy}) + g_{3,Q} Q^2 u_{zz}. 
\end{align}
Thus, below the corresponding critical concentrations, where $L \propto |x'-x'_{c,L}|^{1/2}$ and $Q \propto |x'-x'_{c,Q}|^{1/2}$, the emergence of these orders modify Vegard's law producing the slope changes that can be seen in Fig. \ref{fig:m_ac.vs.x} (bottom panel). Remarkably, the strength of the purely structural and magnetostructural couplings, $g_{i,Q}$ and $g_{i,L}$ respectively, turn out to be comparable in La$_{2-x}$Ba$_x$NiO$_4$. 
}

\begin{figure*}[t!]
\flushleft 
{\hspace{10em} La$_{2}$NiO$_4$ 
\hspace{24em} La$_{1.5}$Ba$_{0.5}$NiO$_4$}\\ \vspace{0.2cm}
    \begin{minipage}[c]{0.49\textwidth}
        \centering
        \includegraphics[width=.99\textwidth]{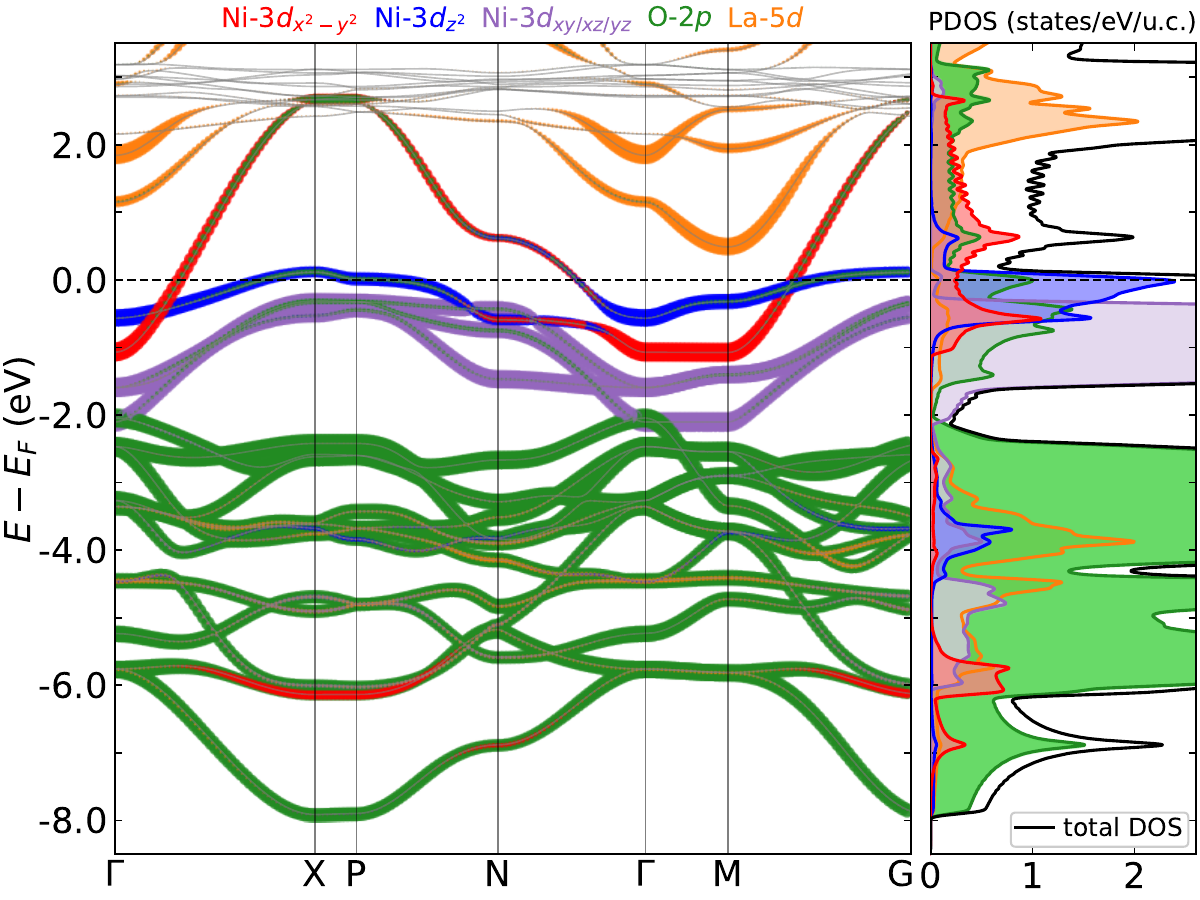} \\
        \includegraphics[width=0.35\textwidth]{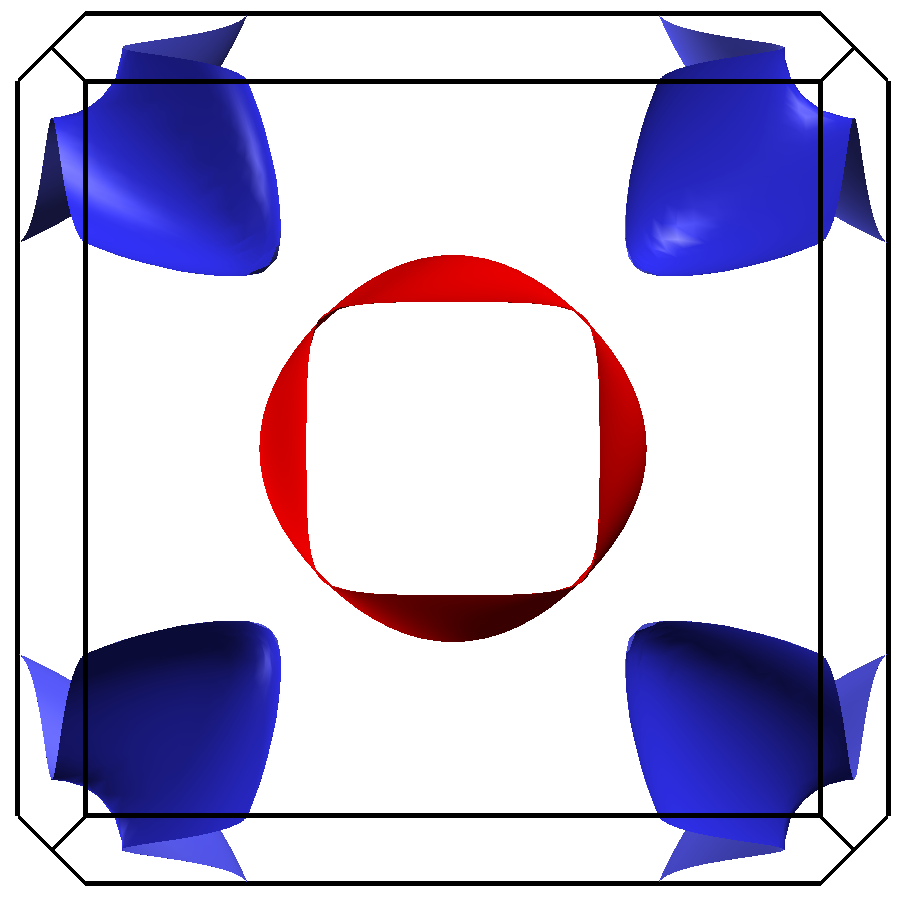} 
        \hspace{1em}
        \includegraphics[width=0.35\textwidth]{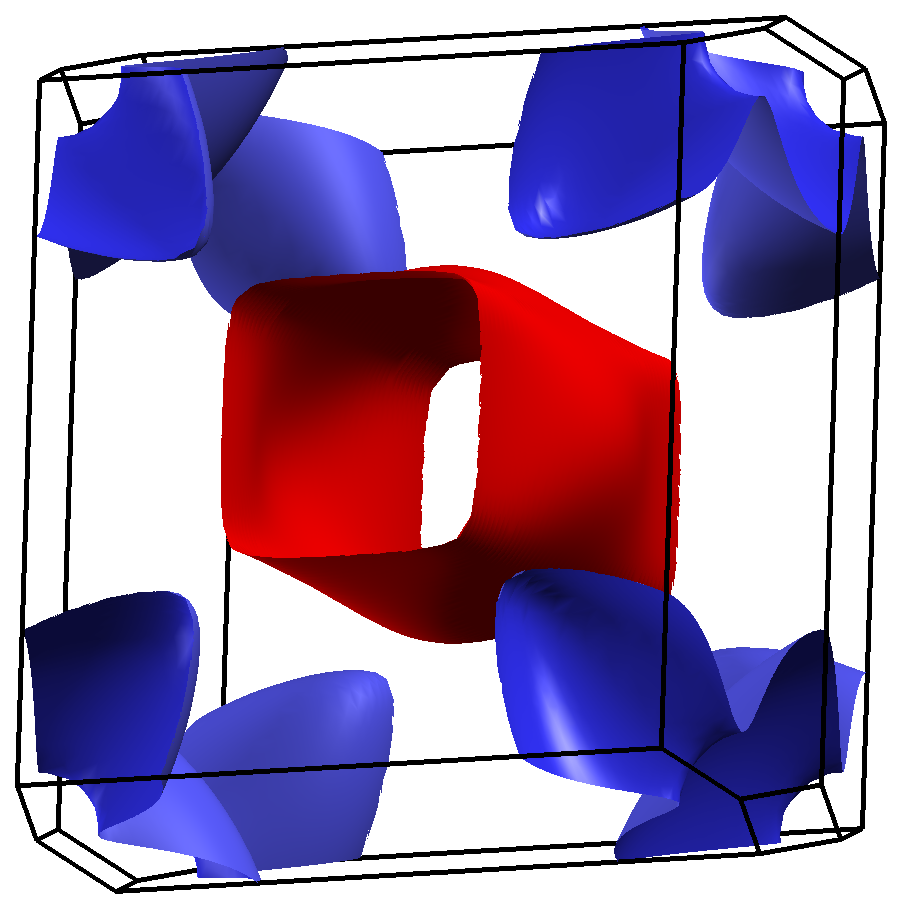} 
    \end{minipage}        
    \hfill 
    \begin{minipage}[c]{0.49\textwidth}
        \centering
        \includegraphics[width=.99\textwidth]{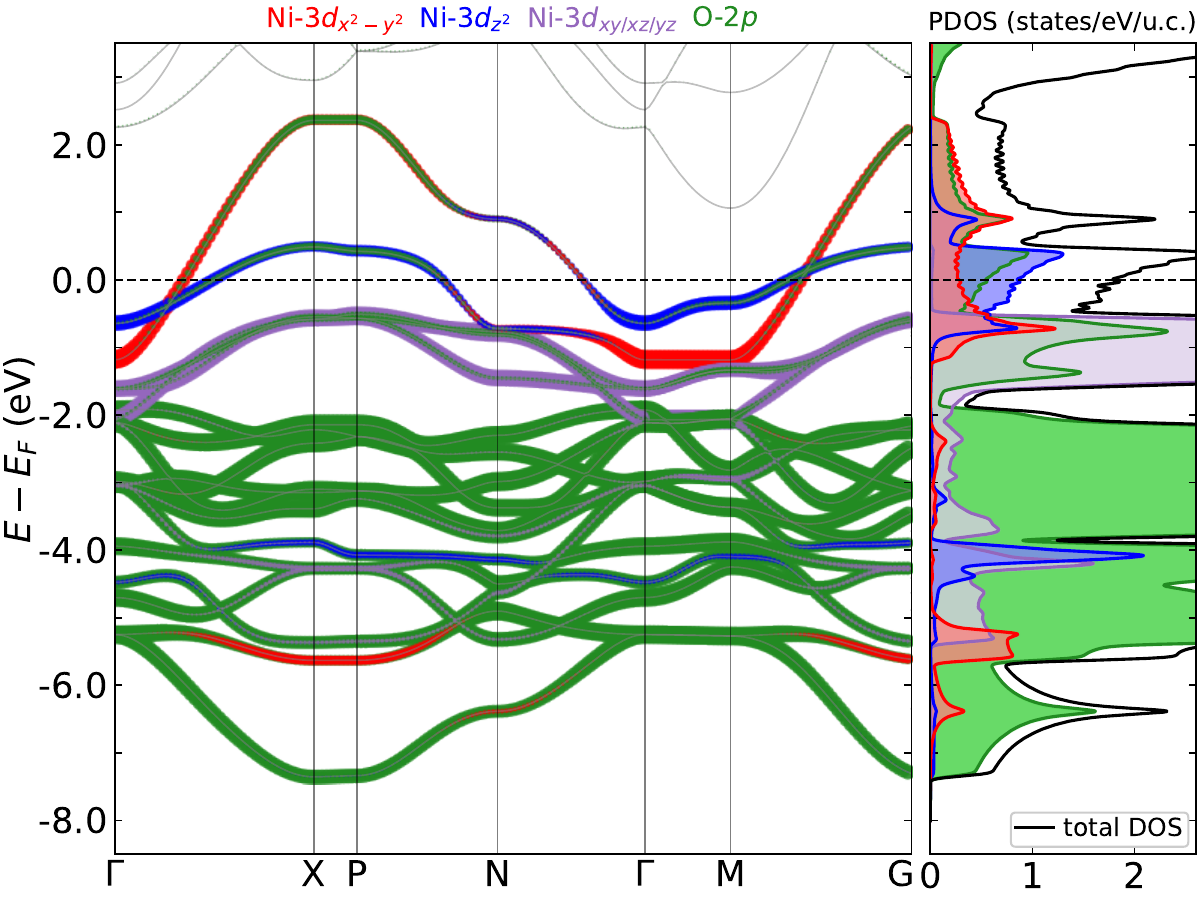}\\
        \includegraphics[width=0.35\textwidth]{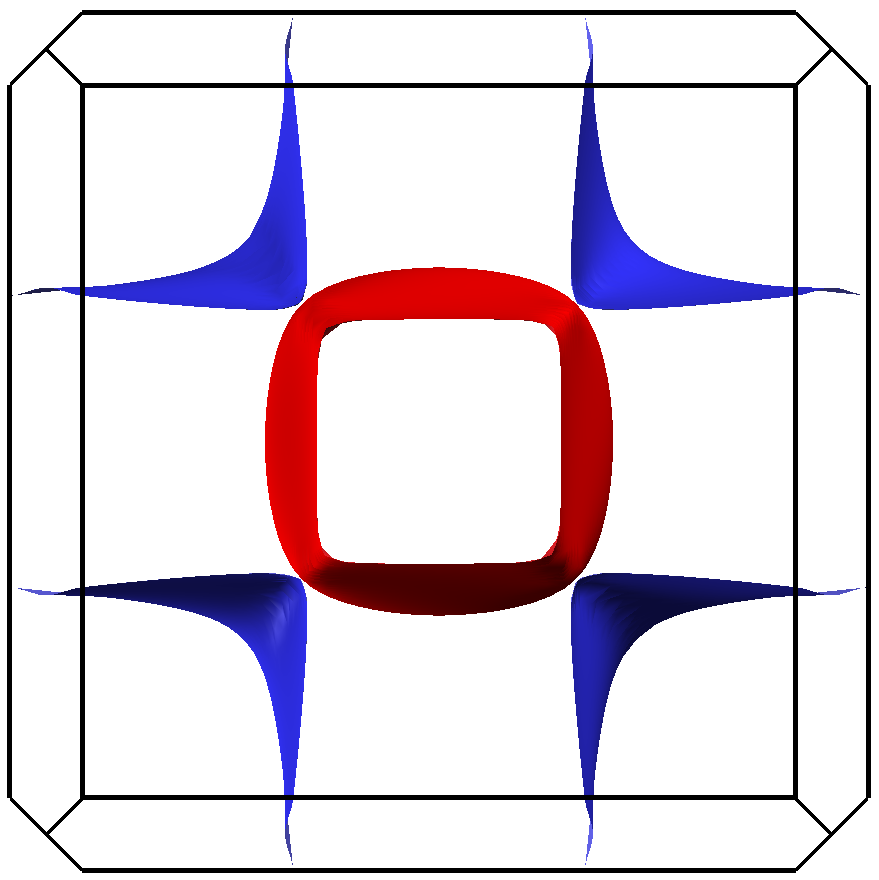} 
        \hspace{1em} 
        \includegraphics[width=0.35\textwidth]{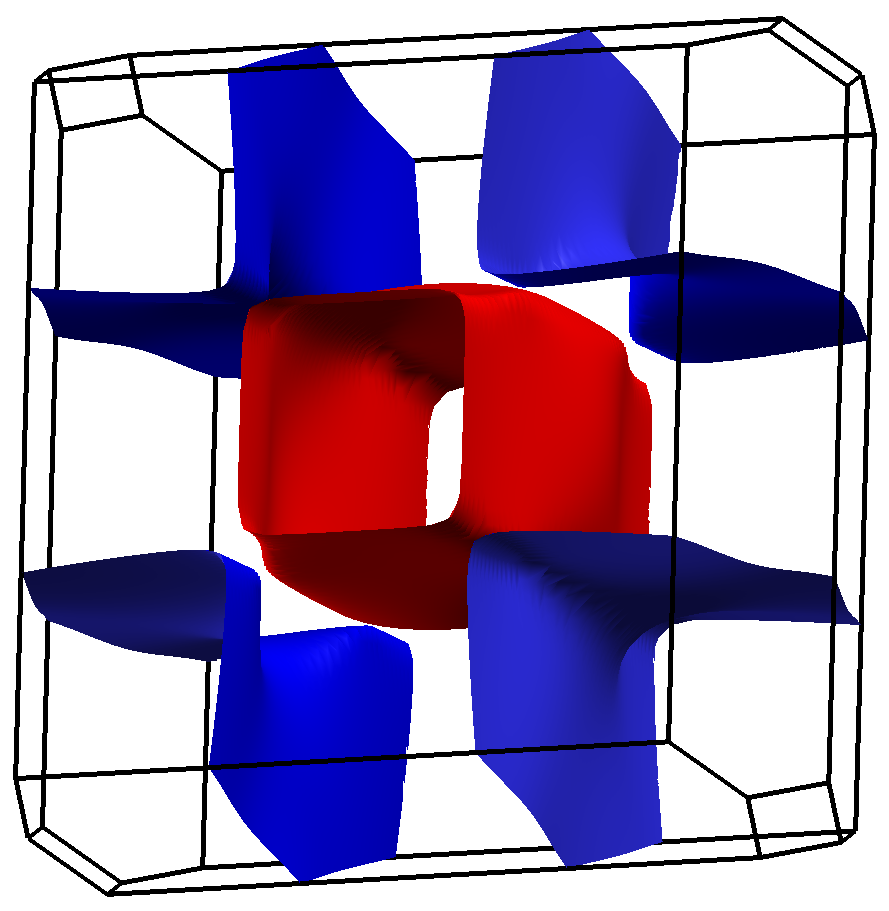} 
    \end{minipage}      
  
\caption{Electronic structures and Fermi surfaces of La$_2$NiO$_4$ in its $I4/mmm$ phase (left) and La$_{1.5}$Ba$_{0.5}$NiO$_4$ (right), both at ambient pressure. 
Note that, while the blue $d_{z^2}$ flat bands yield the blue Fermi surface sheets, 
there is a $k$-dependent mixing between the $d_{x^2-y^2}$ and $d_{z^2}$ states that yield the red Fermi surface sheet. In other words, there is no 100\% correspondence between the colors of the Fermi-surface sheets and their orbital character. 
}
    \label{fig:bands}
\end{figure*}

\vspace{1ex}
\noindent{\bf Electronic structure.} \\
\noindent 
Next, we discuss the evolution of the electronic structure of these systems. The parent compound La$_2$NiO$_4$ is formally associated with a Ni $d^8$ configuration. 
In overall, the calculated electronic structure shows only moderate modifications as a function of pressure, with rather small changes associated with the ``tetragonalization'' itself. 
Thus, we consider the tetragonal $I4/mmm$ structure at ambient pressure to illustrate the main features of the electronic structure of La$_2$NiO$_4$ in the non-spin polarized case {(see Fig. \ref{fig:bandsBmab} for a comparison between $I4/mmm$ and $Bmab$)}. 
The calculated electronic structure is shown in Fig. \ref{fig:bands} (see Fig. \ref{fig:bands8GPa} for the electronic structure at 8~GPa). 
The Ni-$3d$ derived states can be separated according to their projections into $e_{g}=d_{x^2-y^2/z^2}$ and $t_{2g}=d_{xy/xz/yz}$ states, which are partially filled and completely occupied respectively. 
Further, the bandwidth of the Ni-$d_{x^2 -y^2}$ states is $\sim 3.5$~eV, while the Ni-$d_{z^2}$ ones produce a flat-band feature that dominates the DOS at the Fermi level. 
From the projected DOS, we find the occupancies $n_{d_{x^2-y^2}}=1.03$, $n_{d_{z^2}} = 1.60$ and $n_{t_{2g}} = 5.84$. 
The difference from the nominal $d^8$ configuration can be ascribed to the hybridization with the O-2$p$ states. 
In fact, the charge transfer energy is $\sim$~3~eV, which is a typical cuprate-like value similar to that of the superconducting bilayer and trilayer nickelates. 
When it comes to the Fermi surface, it is formed of two sheets of hybridized Ni-$e_g$ states. However, compared to the bilayer and trilayer cases, the $d_{z^2}$ sheet is considerably more dispersive along $k_z$. 
Also, including electronic correlations at the $GW$ level results in a self-doping effect from a La-5$d$ state (see Fig. \ref{fig:GW} in Appendix). 
We note that, although its detailed analysis is beyond the scope of the present work, a similar situation takes place in the bilayer and infinite-layer cases  \cite{olevano20,deVaulx25}.

We now turn to the electronic structure of La$_{1.5}$Ba$_{0.5}$NiO$_4$, where the nominal electronic configuration changes to $d^{7.5}$. 
The calculated electronic structure is illustrated in Fig. \ref{fig:bands} {(see Fig. \ref{fig:bandsBmab} for a comparison with La$_2$NiO$_4$ using the $Bmab$ structure of the latter)}. 
As we see, hole doping with Ba is mainly absorbed by the Ni-$e_g$ manifold, which does not occur in a simple rigid-band-shift fashion. 
Specifically, the Ni-$3d_{x^2-y^2}$ bandwidth decreases while the Ni-$3d_{z^2}$ one increases. 
The occupancies of the different Ni-$3d$ orbitals become $n_{d_{x^2-y^2}}=1.17$, $n_{d_{z^2}} = 1.27$ and $n_{{t_{2g}}} = 5.91$.  
Further, we find that the Ni-$3d_{z^2}$ states still yield the main contribution to the DOS at the Fermi level, 
followed by the O-$2p$ and the Ni-$3d_{x^2-y^2}$ ones. 
In fact, the mixing between Ni-$e_g$ and O-$2p$ states increases as the charge-transfer energy reduces to $\sim$~1.8~eV. 
Further, Fermi surface retains its topology and main qualitative features. 
Specifically, the Fermi-surface sheet associated with the Ni-$3d_{z^2}$ flatband remains present and, despite the larger size of the Ba atom, there is no self-doping effect from the La/Ba-5$d$ states. 
Interestingly, the Ni-$d_{z^2}$ states eventually form a 2D Fermi-surface sheet {resembling} that of the superconducting bilayer nickelate (see e.g \cite{deVaulx25}).

\vspace{1ex}
\noindent{\bf Magnetism.} \\
\noindent 
We now analyze {in more detail the} tendency towards magnetic order of these single-layer compounds {by performing} LDA+$U$ calculations. For this, we consider the following configurations: ferromagnetic (FM), A-type antiferromagnetic (A-AFM) where the first-neighbor {Ni} spins are parallel in-plane and anti-parallel out-of-plane, and G-type antiferromagnetic (AFM) with antiparallel first-neighbor {Ni} spins. Note that C-AFM {with antiparallel first-neighbor Ni-spins in-plane and parallel out-of-plane} is equivalent to G-AFM in the tetragonal $I4/mmm$ {Ruddlesden-Popper} structure which we will consider hereafter. 
In addition, we considered in-plane $(0,\pi)$- stripe AFM order but we were unable to find stable solutions for this configuration.  

We start with La$_{2}$NiO$_4$ as the reference compound. In this case, the G-AFM state minimizes the energy compared to non-spin polarized state as well as to the A-AFM and FM configurations. The energy difference depends on the value of the Hubbard $U$ parameter as shown in Fig. \ref{fig:EvsU-La2}. For $U=0$ and ambient pressure, the system remains metallic and the calculated magnetic moment of the Ni atom in the G-AFM state is significantly lower than its experimental value (specifically, 0.64~$\mu_B$ vs. 1.68~$\mu_B$ \cite{rodriguez-carvajal91}). 
For $U \geq 1$~eV and the G-AFM configuration, a gap opens in the DOS at the Fermi level thus yielding insulating behavior as in the experiments. 
Further, the results of constrained RPA calculations for the bilayer case suggest that $U \simeq 3.5$~eV \cite{werner23-prl} is a realistic value for these systems. In fact, with $U = 3.5$~eV the Ni moment increases to 1.4~$\mu_B$, which better matches the experimentally observed value \cite{rodriguez-carvajal91} (see also \cite{bernardini25} for additional calculations in the experimental $Bmab$ structure). 
At the same time, we observe that the calculated moment varies for the different magnetic configurations (see Fig. \ref{fig:EvsU-La2}). This difference suggests an intermediate situation between fully localized spins and itinerant magnetism, signalling that a mapping to a Heisenberg model could be problematic. 
In fact, unlike the G-AFM state in which a gap opens, the A-AFM and FM states remain metallic for all considered $U$ values.

\begin{figure}[t!]
\centering
\includegraphics[width=.325\textwidth]{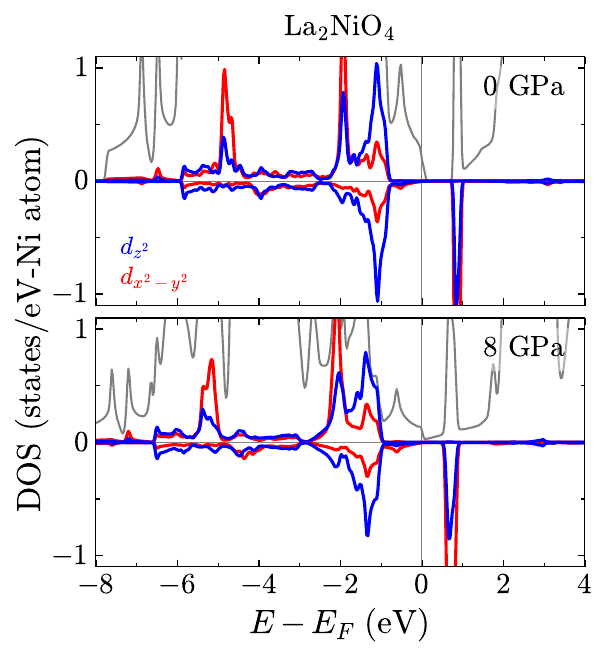}
\caption{Density of states (DOS) projected on the Ni-$d_{z^2}$ (blue) and Ni-$d_{x^2-y^2}$ (red) orbitals in the G-AFM ground state of La$_{2}$NiO$_4$ at ambient pressure and 8 GPa ($U = 3.5$~eV). Spin up/down states are indicated with positive/negative values respectively. The gray curves correspond to the total DOS.
}
\label{fig:DOS-La2}
\end{figure}

Under pressure, in the range from 0 to 8 GPa, we find that the tendency towards magnetic order of La$_{2}$NiO$_4$ remains strong, with just a slight decrease in overall magnetization energy and small changes in the relative energy of the studied magnetic configurations (see Fig. \ref{fig:EvsU-La2}). 
Furthermore, for $U = 3.5$~eV, we find an insulator-to-metal transition for the G-AFM state at $\sim 4$~GPa (below the orthorhombic-to-tetragonal structural transition). 
This circumstance is illustrated in Fig. \ref{fig:DOS-La2}, which shows the corresponding DOS at 0~GPa and 8~GPa.

\begin{figure}[t!]
\centering
\includegraphics[width=.375\textwidth]{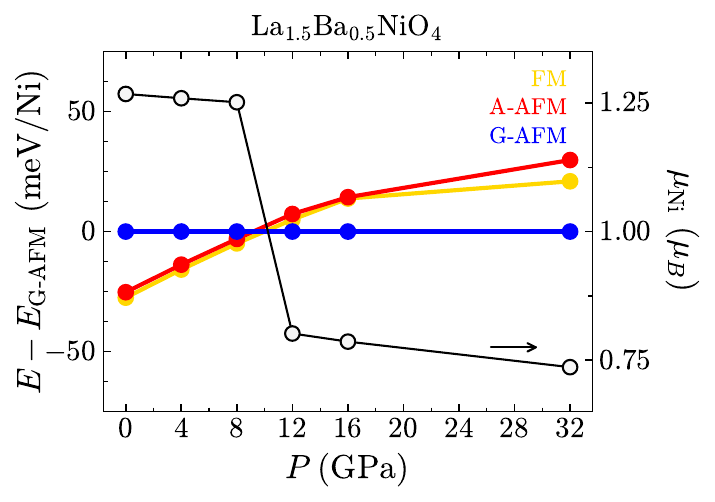}
\caption{Relative energy and Ni magnetic moment as a function of pressure calculated for La$_{1.5}$Ba$_{0.5}$NiO$_4$ ($U = 3.5$~eV).
}
\label{fig:EvsP-LaBa}
\end{figure}
\begin{figure}[t!]
\centering
\includegraphics[width=.485\textwidth]{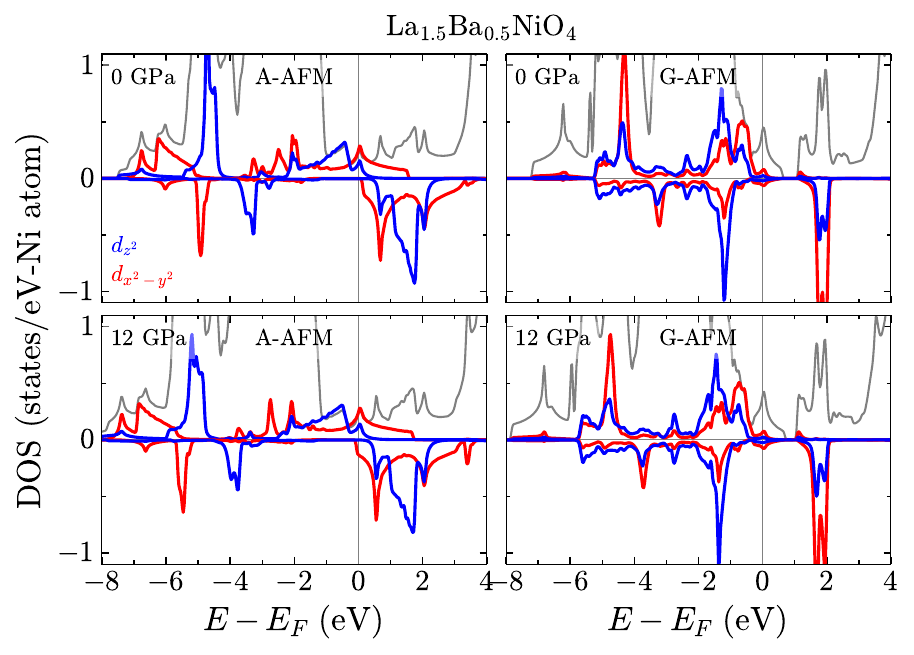}
\caption{Density of states (DOS) projected on the Ni-$d_{z^2}$ (blue) and Ni-$d_{x^2-y^2}$ (red) orbitals in the A-AFM and G-AFM configurations of La$_{1.5}$Ba$_{0.5}$NiO$_4$ at ambient pressure and 12 GPa ($U = 3.5$~eV). Spin up/down states are indicated with positive/negative values respectively. The gray curves correspond to the total DOS. 
}
\label{fig:DOS-LaBa}
\end{figure}

Finally, we consider La$_{1.5}$Ba$_{0.5}$NiO$_4$. 
In this case, we obtain reduced magnetization energies together with an intriguing competition between the different magnetic configurations as illustrated in Figs. \ref{fig:EvsP-LaBa} and \ref{fig:EvsU-LaBa}. 
Notably, we find that the FM and A-AFM solutions are essentially degenerate. This is a manifestation of the enhanced low-dimensional character of the system---also reflected in the shape of the Fermi surface---implying an effective decoupling in the out-of-plane direction even under pressure. 
Moreover, the relative energy with respect to the G-AFM configuration turns out to be significantly reduced also, as we find it always below $\sim$~50~meV/Ni (see Fig. \ref{fig:EvsP-LaBa}).

Interestingly, this competition remains as a function of both $U$ and pressure.
For $U \lesssim 3.5$~eV, the configuration that minimizes the energy corresponds to the G-AFM state. This can be seen as reminiscent of the parent compound. 
For $\gtrsim 3.5$~eV, however, the situation is reversed, in the sense that the energy is minimized by FM/A-AFM configurations. 
In addition, for $\gtrsim 3.5$~eV, we find an intriguing crossover from FM/A-AFM to G-AFM as a function of pressure, which is illustrated in Fig. \ref{fig:EvsP-LaBa} (see also Fig. \ref{fig:EvsU-LaBa}). 
This crossover implies a drop in the magnetic moment of the Ni atom of 0.5~$\mu_B$. \\[-2ex]

\noindent{\bf Discussion.} \\ 
\noindent
We now turn to the implications of our results, beginning with the {magnetostructural behavior. Our calculations have revealed a substantial interplay between the octahedral tilts, magnetic order and strain in (La,Ba)$_2$NiO$_4$. 
This interplay is a key factor in setting the critical pressure at which La$_2$NiO$_4$ becomes tetragonal, and further explains the unusual evolution of the lattice parameters with composition observed experimentally, which deviates from Vegard's law. 
} 

{Further, the comparison between La$_2$NiO$_4$ and La$_{1.5}$Ba$_{0.5}$NiO$_4$ reveals the emergence of} 
an intriguing competition between different magnetic configurations. This can lead to magnetic frustration and, consequently, to a paramagnetic-like disordered state. This scenario is {in} fact consistent {with} the glass-like behavior reported experimentally {when the La:Ba ratio is 1:1} \cite{schilling09}.  

{Moreover}, if the correlations are sufficiently strong,  
this competition can appear as a crossover from FM/A-AFM to G-AFM order that, at first glance, resembles the one obtained for the bilayer La$_{3}$Ni$_2$O$_7$ \cite{labollita2024-bilayer}. 
However, a closer inspection of the electronic structure reveals a redistribution of both $d_{x^2-y^2}$ and $d_{z^2}$ states in the case of La$_{1.5}$Ba$_{0.5}$NiO$_4$ (see Fig. \ref{fig:DOS-LaBa}), pointing to a distinct microscopic origin of this crossover. 
Furthermore, the observation that La$_2$NiO$_4$ becomes metallic under either physical or chemical pressure suggests a possible modification of the magnetic-coupling mechanisms, evolving from superexchange toward direct exchange.

Regarding superconductivity, early theoretical studies emphasized the crucial role of the multilayer structure for superconductivity in the Ruddlesden-Popper nickelates,
drawing direct analogies with the bilayer Hubbard model where $s_\pm$-wave pairing arises near half-filling for sufficiently large interlayer coupling (see, e.g., \cite{kuroki17-prb,kuroki24-prl}). 
{Subsequent works have also suggested the possibility of $d$-wave superconductivity} in these systems, with multiband effects playing a different role \cite{wu24-prl,xia25-ncomm,lechermann23-prb,kuroki24-prl,savrasov24-prb,dagotto24,xu25}. 
Our results demonstrate that several key ingredients of this broader picture are present in the single-layer systems and can be tuned through pressure and doping. In particular, Ba substitution in La$_2$NiO$_4$ promotes a metallic fermiology with partial analogies to the superconducting multilayers, alongside a similar tendency toward magnetism where different orders may compete. Notably, the Ni-$d_{z^2}$ flatband supplemented with Ni-$d_{x^2-y^2}$ states can form a strongly two-dimensional Fermi surface, which could potentially support $d$-wave superconducting instabilities. 
At the same time, (La,Ba)$_2$NiO$_4$ lacks the multiband features necessary for multilayer-Hubbard-model-like $s_\pm$-wave superconductivity, making it an interesting case-study system to isolate the fundamental ingredients behind the competition of different superconducting instabilities proposed for the multilayer nickelates {(see also \cite{kuroki20} for a theoretical consideration of superconductivity in La$_2$NiO$_4$)}.\\[-1ex]

\noindent{\bf Conclusions.} \\
\noindent 
We have investigated the effects of pressure and doping on the crystal and electronic structure of the reference single-layer nickelate La$_2$NiO$_4$. 
We have found that the suppression of oxygen-octahedral tilts, driving a structural change to a high-symmetry tetragonal $I4/mmm$ phase, occurs at 
{similar} 
pressures in La$_2$NiO$_4$ compared to its bilayer and trilayer superconducting counterparts. 
Remarkably, this structural change can also be achieved at ambient pressure through partial substitution of La with Ba.
{In this case, we have uncovered a distinct magnetostructural interplay that explains the deviation from Vegard's law observed experimentally. 
}
Beyond stabilizing the tetragonal structure {and eventually suppressing magnetism}, we have {also} shown that Ba doping induces subtle but consequential modifications to the electronic structure, shifting it toward a nominal $d^{7.5}$ configuration similar to the superconducting bilayer compounds, as a result of which the two-dimensional character of the Ni-$d_{z^2}$ Fermi-surface sheet is enhanced and there appears an intriguing competition between different magnetic configurations.

Our results thus position divalent-atom substituted single-layer nickelates like  (La,Ba)$_2$NiO$_4$ as 
{valuable} model systems, {not only for probing strong magnetostructural interplays, but also for mirroring key electronic} 
features of the superconducting Ruddlesden-Popper nickelates, while lacking the full multiband character of the bi- and trilayer compounds. 
Precisely because of this latter simplification, they offer a valuable platform for disentangling the respective roles of structural symmetry, dimensionality, and electronic configuration. 
Our findings thus clarify essential aspects of La$_2$NiO$_4$ and provide a foundation for future studies aimed at isolating the key ingredients behind superconductivity in the nickelates.

\vspace{1ex}
\noindent{\it Acknowledgments.} 
We acknowledge the LANEF Chair of Excellence program for funding. 
F. B. acknowledge partial support from Italian Ministry of University and Research MUR, financed by the European Union - Next Generation EU, within the PRIN 2022 call, contract n. 2022M3WXE7. Computational resources were provided by the GRICAD supercomputing center of Université Grenoble Alpes and GENCI Grant No. 2022-AD01091394. We thank J. Even and C. Katan for useful discussions.

\bibliography{bib.bib}

\clearpage

\onecolumngrid
\begin{center}
    {\bf Appendix}
\end{center}

\renewcommand{\thefigure}{A\arabic{figure}}
\renewcommand{\thetable}{A\arabic{table}}
\setcounter{figure}{0}
\setcounter{table}{0}

\begin{figure}[h!]
\centering
\includegraphics[height=.22\textwidth]{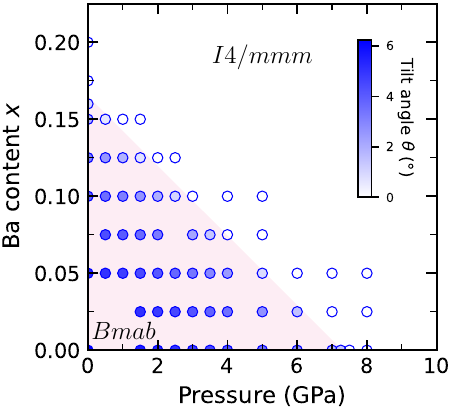}
\caption{{Composition versus pressure phase diagram calculated for La$_{2-x}$Ba$_x$NiO$_4$ in the non-spin polarized case, where the shade of the circles correspond to the calculated the tilt angle.}
}\label{fig:PhD}
\end{figure}

\begin{figure*}[h!]
        \centering
        \includegraphics[height=.225\textwidth]{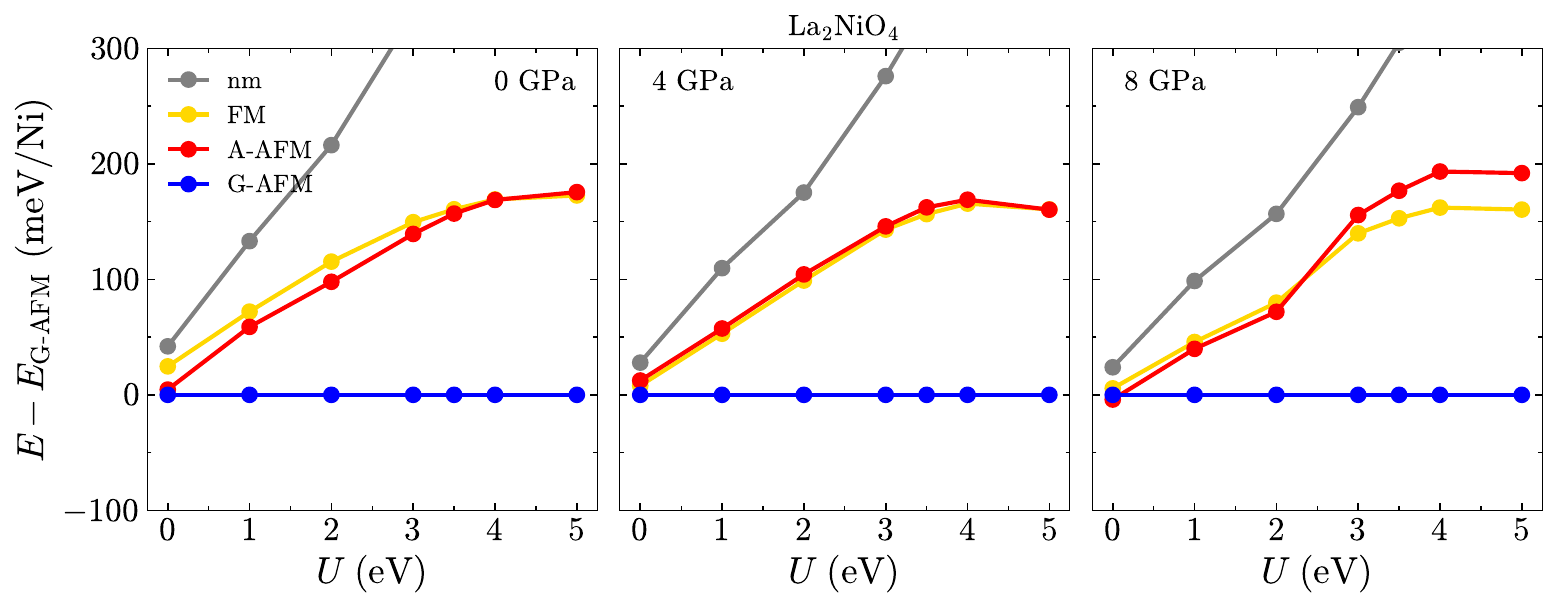}
        \hspace{2em}
        \includegraphics[height=.225\textwidth]{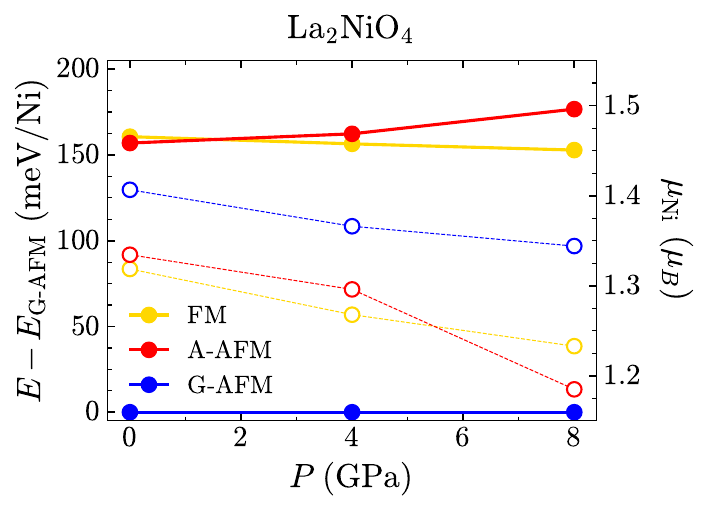}
        \vspace{2em}
    \caption{Relative energy of the different magnetic configurations of La$_2$NiO$_4$ calculated as a function of the Hubbard $U$ parameter for different pressures. 
    }
    \label{fig:EvsU-La2}
\end{figure*}

\begin{figure*}[h!]
    \includegraphics[width=.995\textwidth]{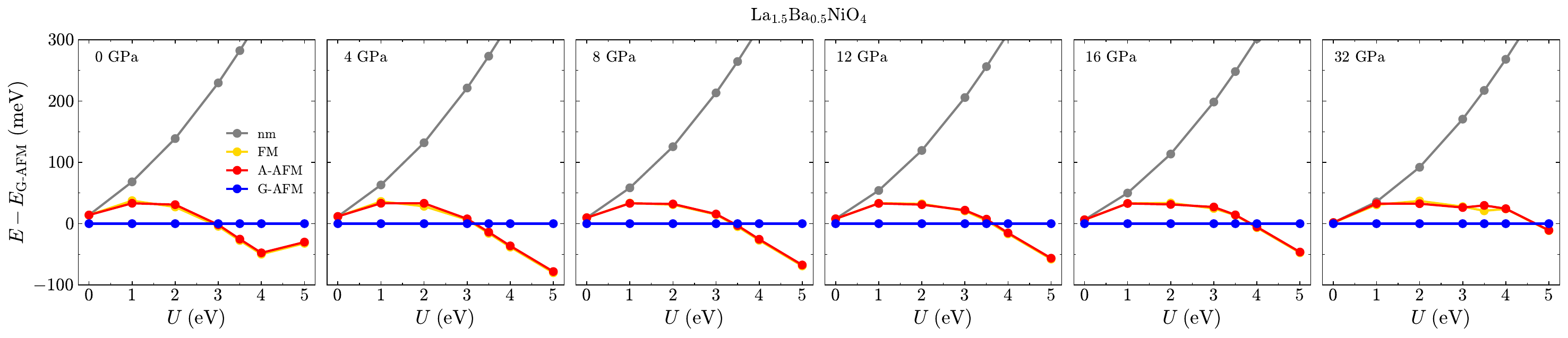}
    \caption{Relative energy of the different configurations of La$_{1.5}$Ba$_{0.5}$NiO$_4$ calculated as a function of the Hubbard $U$ parameter for different pressures. 
    }
    \label{fig:EvsU-LaBa}
\end{figure*}

\begin{figure*}[h!]
\flushleft\hspace{16.5em} La$_2$NiO$_4$ \hspace{23em} La$_{1.5}$Ba$_{0.5}$NiO$_4$ \\
\vspace{-1em}
\flushleft\hspace{7em} $Bmab$ \hspace{16em} $I4/mmm$ \hspace{14em} $I4/mmm$ \\
\vspace{1ex}
\centering
\includegraphics[width=0.3\textwidth]{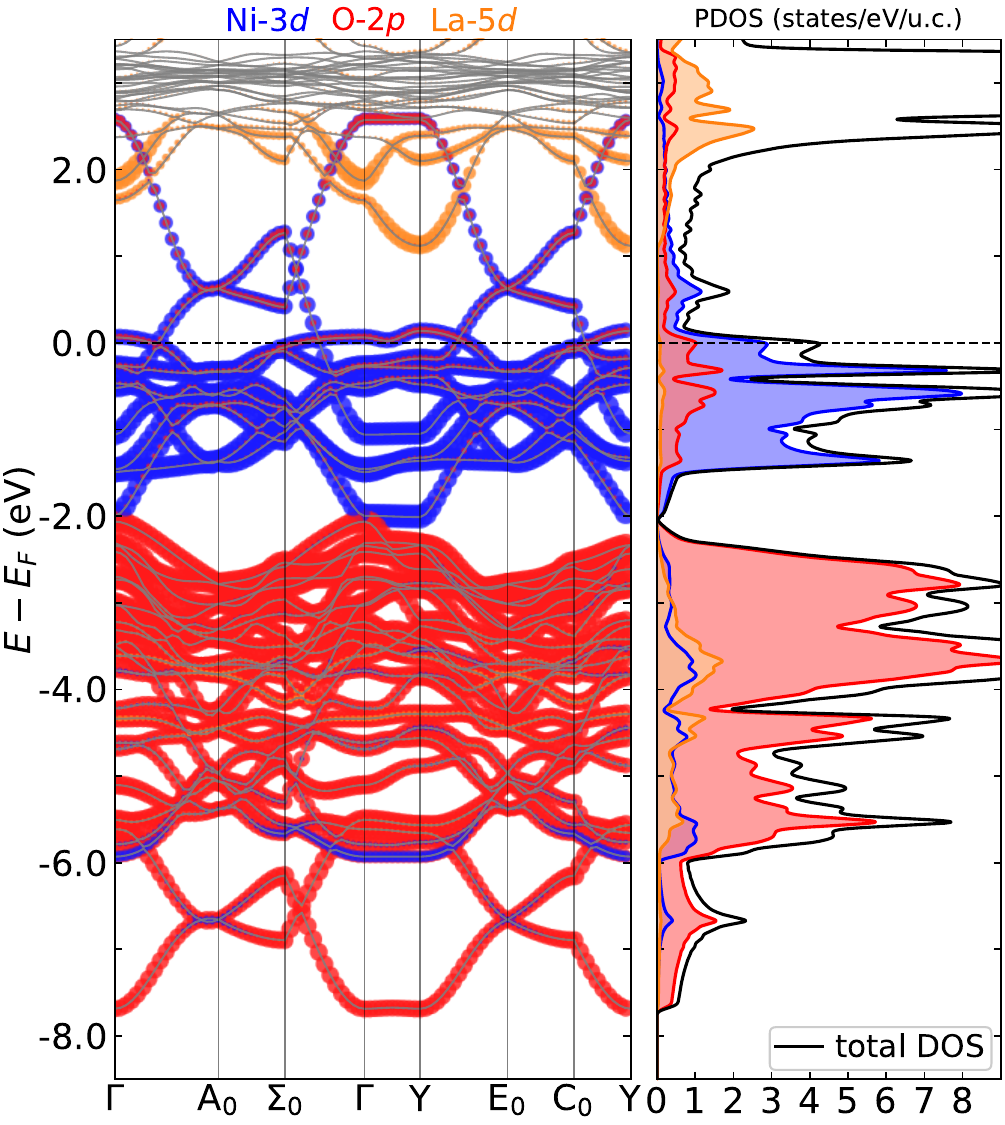}
\hspace{2em}
\includegraphics[width=.3\textwidth]{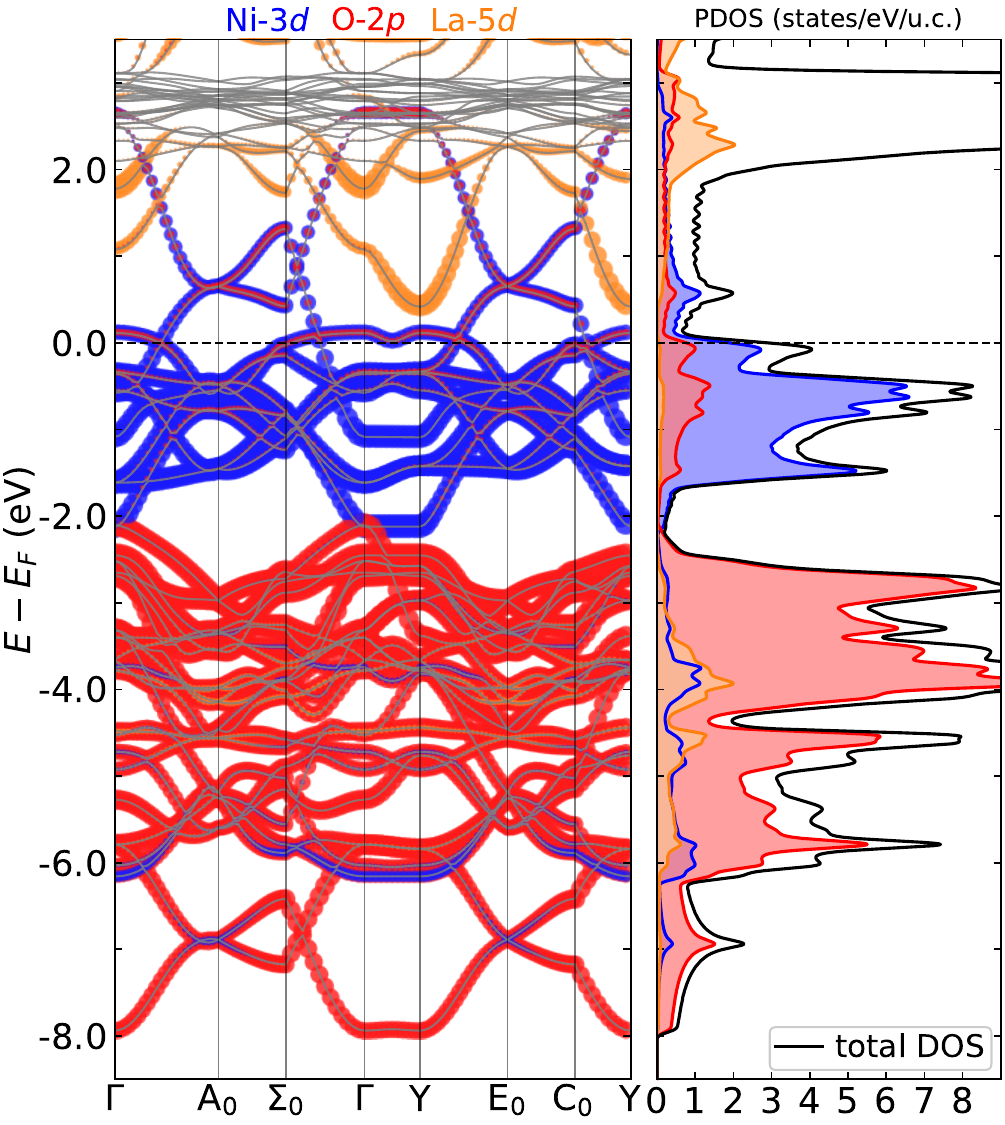}
\hspace{2em}
\includegraphics[width=0.3\textwidth]{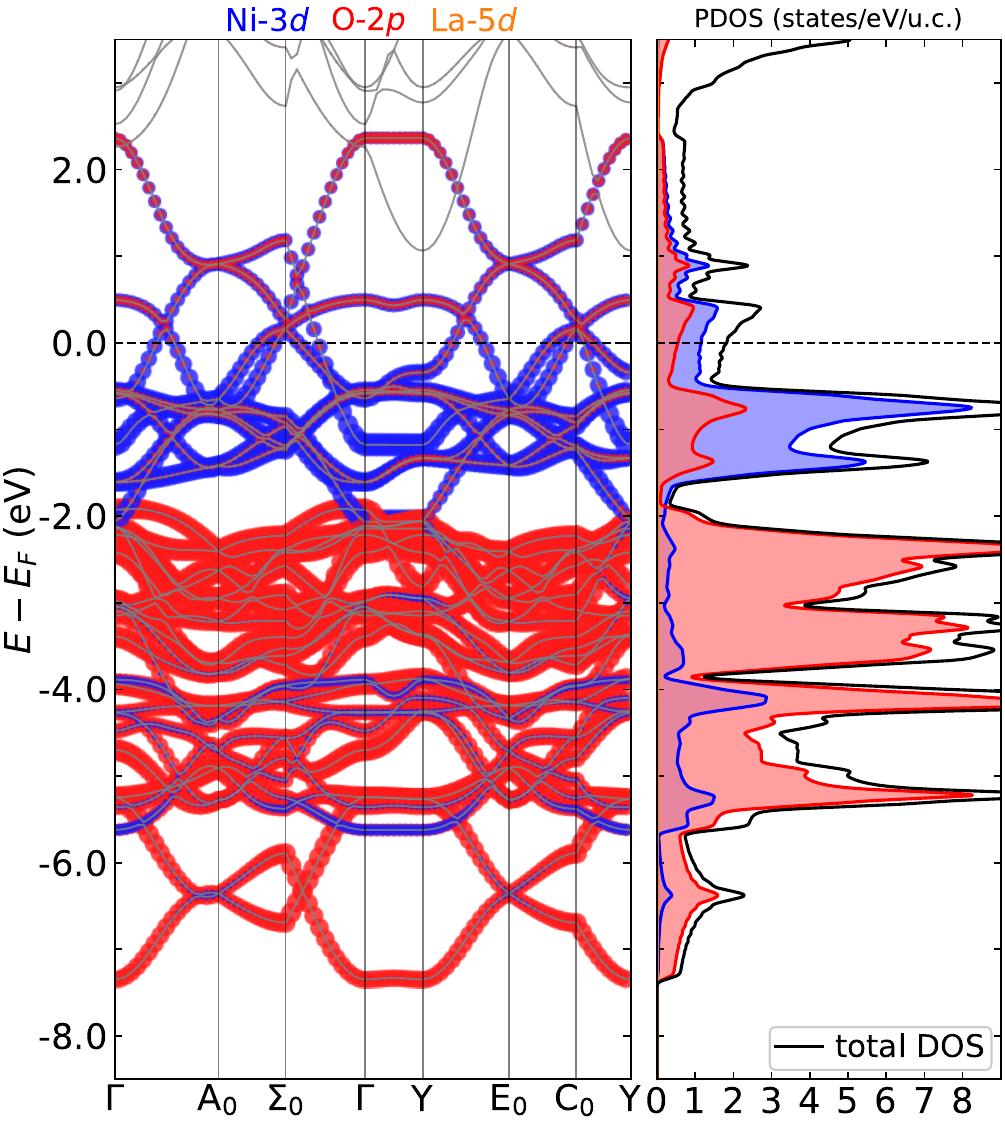}
\\ 
    \caption{
Comparison between the non-spin polarized band plot and DOS of the parent compound La$_2$NiO$_4$ in its $Bmab$ and $I4/mmm$ phases and La$_{1.5}$Ba$_{0.5}$NiO$_4$ (using analogous $Bmab$ supercells for the band plots). 
}
\label{fig:bandsBmab}
\end{figure*}

\begin{figure*}[h!]
  \centering
  \begin{minipage}[c]{0.63\textwidth}
    \centering
    \includegraphics[width=\textwidth]{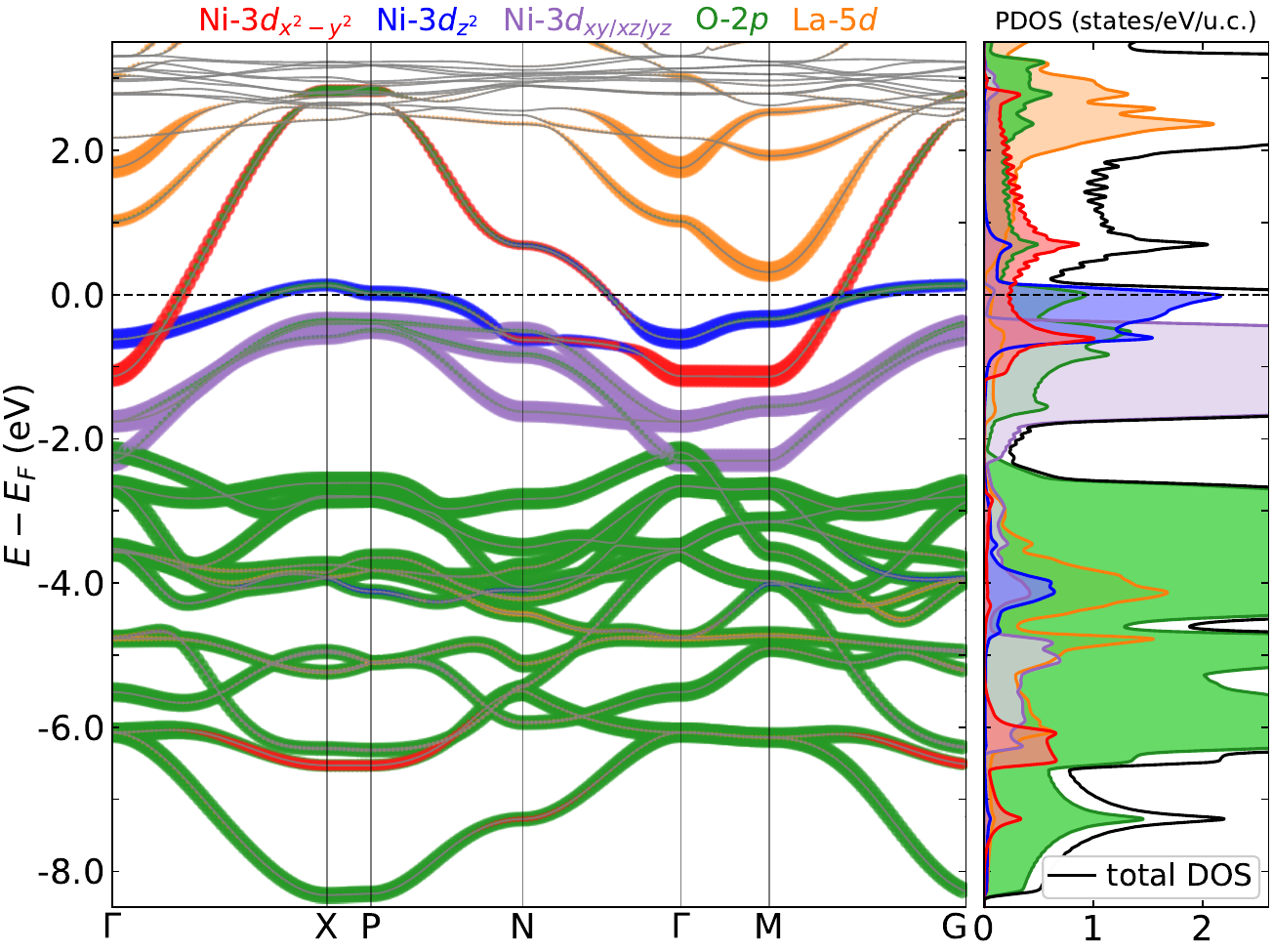}
  \end{minipage}
  \hfill 
  \begin{minipage}[c]{0.35\textwidth}
    \centering
    \includegraphics[width=0.55\textwidth]{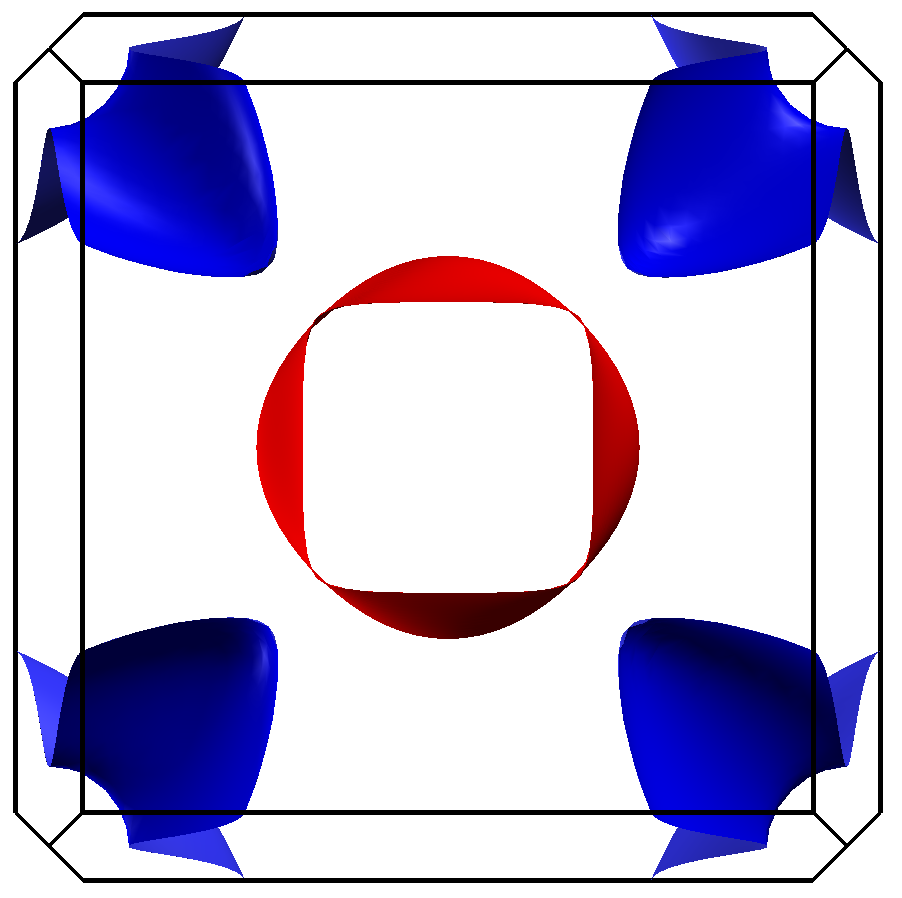}\\ \vspace{0.3cm}
    \includegraphics[width=0.55\textwidth]{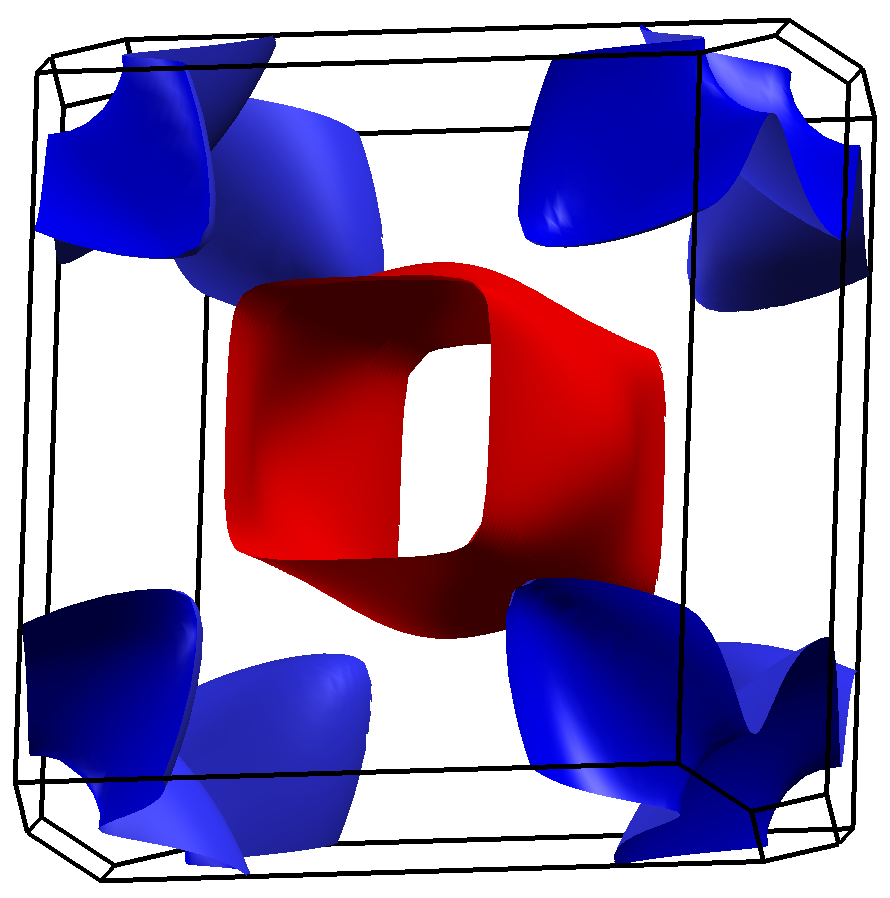}
    \end{minipage}
    \caption{
Electronic structure, PDOS and Fermi surface of La$_2$NiO$_4$ at 8 GPa, to compare with Fig.~\ref{fig:bands} which is at ambient pressure. The main difference is only in the dispersion which slightly increases, while the Fermi surface is almost unaffected.
}
\label{fig:bands8GPa}
\end{figure*}

\begin{figure*}[h!]
  \centering
  \begin{minipage}[c]{0.60\textwidth}
    \centering
    \includegraphics[width=\textwidth]{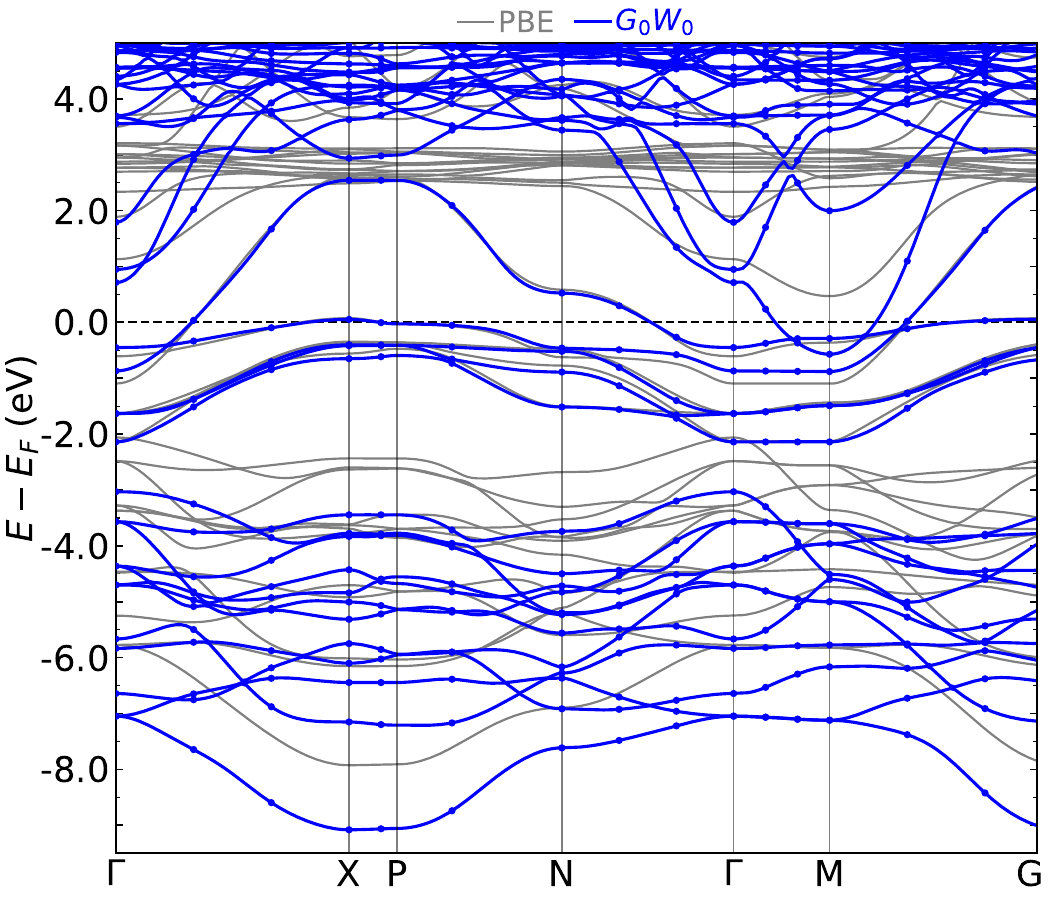}
  \end{minipage}
  \hfill 
  \begin{minipage}[c]{0.38\textwidth}
    \centering
    \begin{tabular}{c}
        \includegraphics[width=0.49\textwidth]{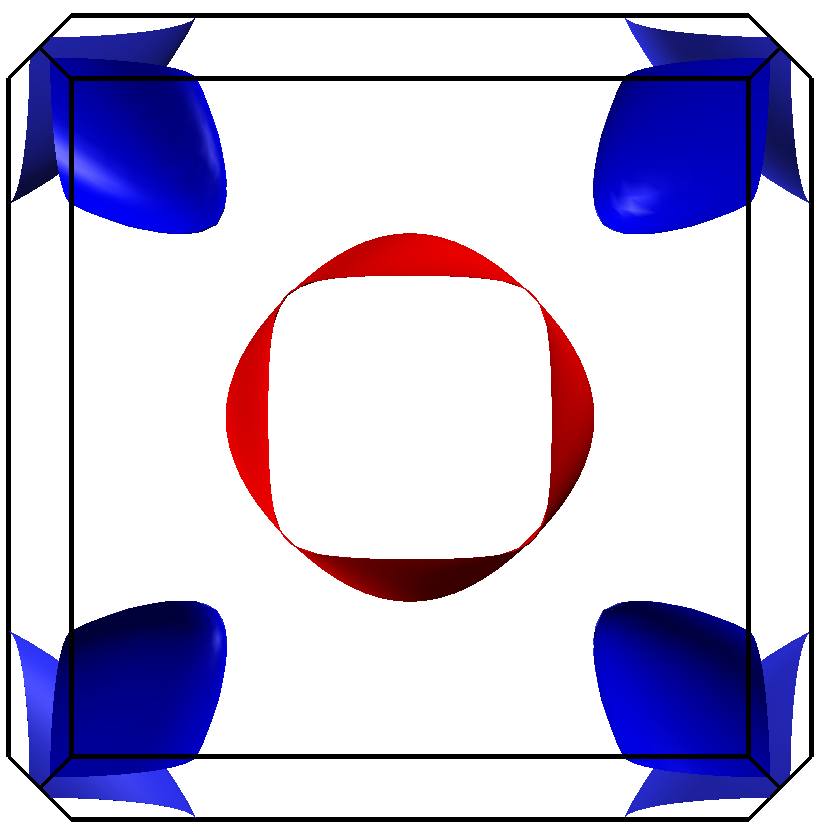}\hfill
        \includegraphics[width=0.49\textwidth]{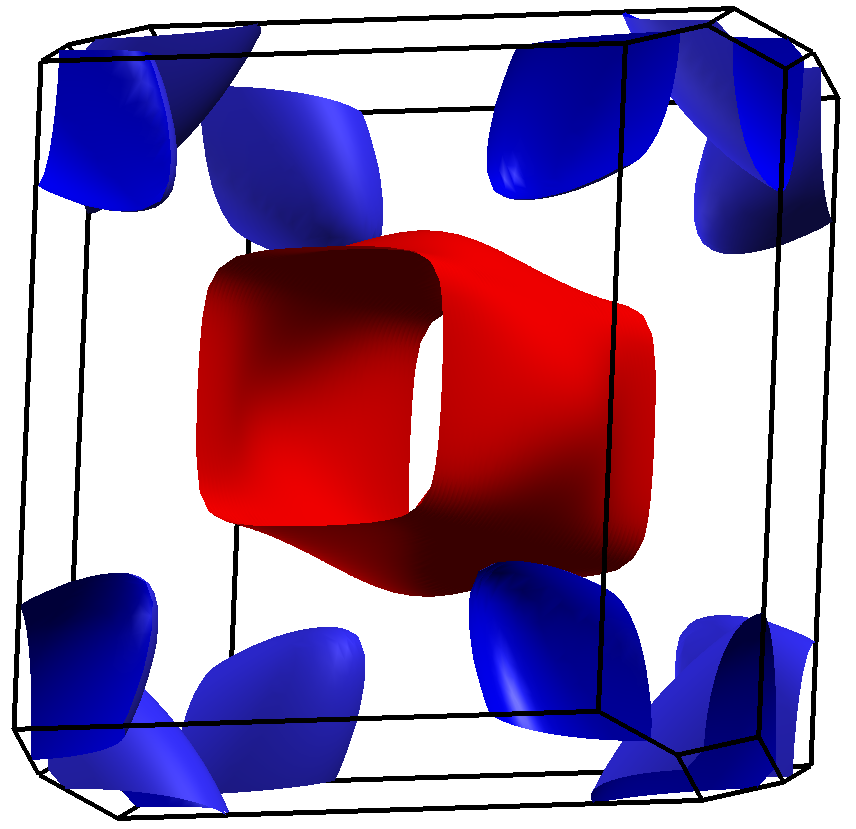}
        \\DFT \vspace{0.7cm}\\
        \includegraphics[width=0.49\textwidth]{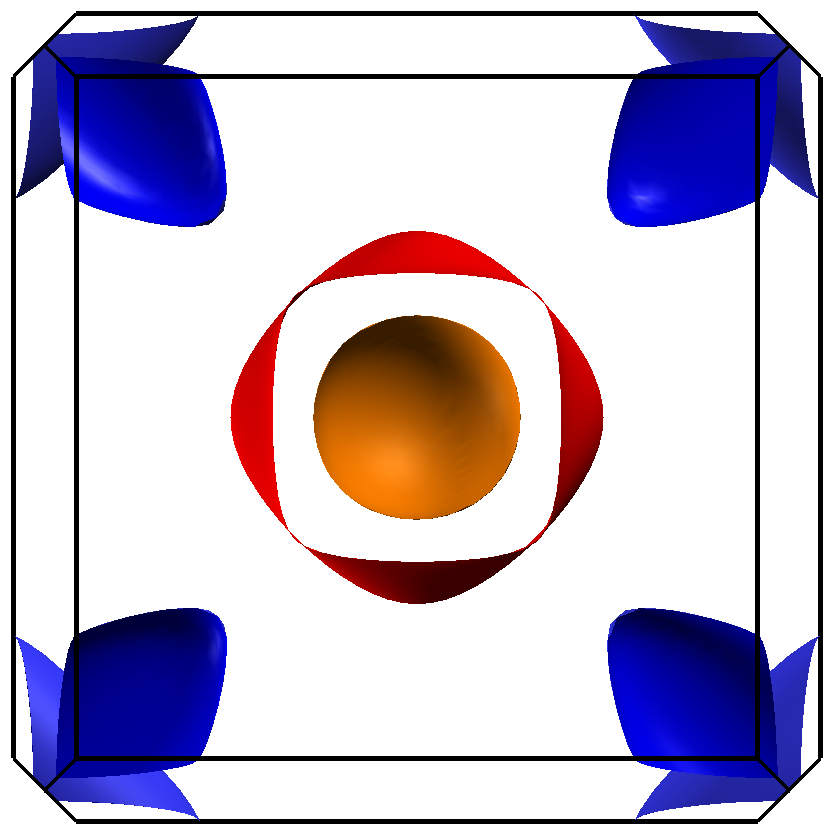} \hfill
        \includegraphics[width=0.49\textwidth]{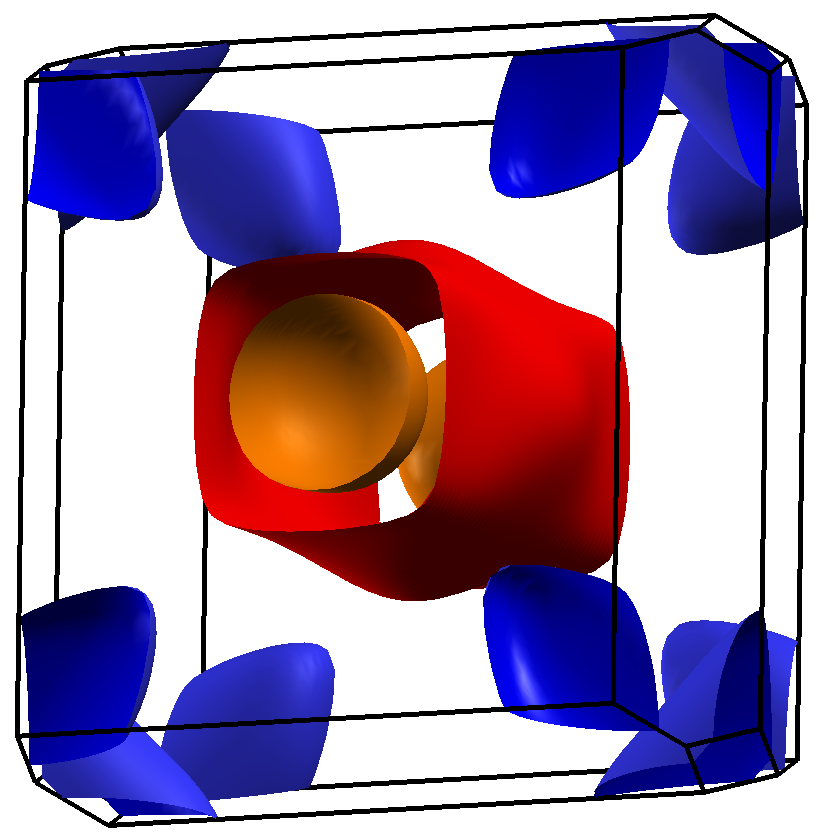}\\$GW$
    \end{tabular}
    \end{minipage}
    \caption{
Comparison between the calculated electronic structure and Fermi surfaces of La$_2$NiO$_4$ at ambient pressure at the DFT level and including further correlations within the $GW$ approximation for the self-energy (dots correspond to quasiparticle eigenenergies on the $6\!\times\!6\!\times\!6$ grid and the solid line to its Wannier interpolation). 
In this case, we considered a fully relaxed $I4/mmm$ tetragonal structure whose lattice parameters are $a = 3.794$ {\AA} and $c= 13.039${\AA}.\\[1em]
For the $GW$ calculation, we used the plane-wave-based code \textsc{Abinit} \cite{Abinit}.
We have used a 16$\times$16$\times$16 standard sampling of the Brillouin zone (BZ) and a plane waves (PW) cutoff of 65~Ha for the DFT-PBE starting calculation, as well as a Gaussian smearing of 0.01 Hartree. In the subsequent $GW$ calculations, the BZ sampling has been reduced to 6$\times$6$\times$6, whereas the PW cutoffs have been reduced to 40 Ha for the representation of the wave functions and for the exchange self-energy, and further down to 15 Ha for the correlation self-energy and the screening. 
We included 170 bands for the calculation of the screening and 200 for the self-energy. We have carried just only one $GW$ iteration (i.e.\ $G_0W_0$) on top of PBE, using a Godby-Needs plasmon-pole model \cite{GodbyNeeds89} and a shift of 0.1~eV to avoid poles/divergences. The final $GW$ band plots and Fermi surfaces, were interpolated from the 6$\times$6$\times$6 to a denser $k$-mesh by the \textsc{Wannier90} code using a set of 47 maximally-localized Wannier functions (MLWFs) namely O-$2p$, Ni-$3d$, Ni-$4p$, Ni-$3s$, La-$5d$, La-$4f$ and La-$3s$. All calculations have been done with a Gaussian smearing temperature $T_s$ of 0.01~Ha.\\[1em]
One can remark a difference between the Ni-3$d_{z^2}$ Fermi sheets in the corners of the BZ (blue) with respect to the ones shown on Fig. \ref{fig:bands}. Since the $d_{z^2}$ band is very flat, the corresponding Fermi sheet is extremely sensible to relatively small changes. 
The other Fermi sheets, namely mainly Ni-3$d_{x^2-y^2}$ (red) and La-5$d_{x^2-y^2}$ (orange) are more robust in this respect.
The self-doping induced by the $GW$ lowering of the La-5$d_{x^2-y^2}$ band is quite important. We note that the situation is analogous in the bilayer \cite{deVaulx25} where the self-doping occurs beyond $\sim$30 GPa, and to a certain extent in the infinite layer. In the latter, the self-doping by the rare-earth band is already present at the DFT level, at different $k$-points of the BZ and with different orbitals involved, namely a mixture of $d_{z^2}$ and $d_{xy}$ states rather than $d_{x^2-y^2}$.
}
\label{fig:GW}
\end{figure*}

\end{document}